\documentclass[aps,prd,preprint,showpacs,12pt]{revtex4}
\usepackage{amssymb}
\usepackage{epsfig}
\usepackage{bm}

\newcommand\Tr{\mathop{\rm Tr}\nolimits}
\newcommand\Ln{\mathop{\rm Ln}\nolimits}
\newcommand\arccot{\mathop{\rm arccot}\nolimits}

\begin{document}

\date{December 3, 2002}

\title{
On the evaluation of the improvement parameter\\
in the lattice Hamiltonian approach to critical phenomena}

\author{Massimo Campostrini}%
\email{Massimo.Campostrini@df.unipi.it}%
\affiliation{INFN, Sezione di Pisa, and
Dipartimento di Fisica ``Enrico Fermi'' dell'Universit\`a di Pisa,
Via Buonarroti 2, I-56125 Pisa, Italy}

\author{Pietro Parruccini}%
\email{Pietro.Parruccini@df.unipi.it}%
\affiliation{INFN, Sezione di Pisa, and
Dipartimento di Fisica ``Enrico Fermi'' dell'Universit\`a di Pisa,
Via Buonarroti 2, I-56125 Pisa, Italy}

\author{Paolo Rossi}%
\email{Paolo.Rossi@df.unipi.it}%
\affiliation{INFN, Sezione di Pisa, and
Dipartimento di Fisica ``Enrico Fermi'' dell'Universit\`a di Pisa,
Via Buonarroti 2, I-56125 Pisa, Italy}

\begin{abstract}
In lattice Hamiltonian systems with a quartic coupling $\gamma$, a
critical value $\gamma^*$ may exist such that, when $\gamma=\gamma^*$,
the leading irrelevant operator decouples from the Hamiltonian and the
leading nonscaling contribution to renormalization-group invariant
physical quantities (evaluated in the critical region) vanishes.  The
$1/N$ expansion technique is applied to the evaluation of $\gamma^*$
for the lattice Hamiltonian of vector spin models with $O(N)$
symmetry.  As a byproduct, systematic asymptotic expansions for the
relevant lattice massive one-loop integrals are obtained.
\end{abstract}

\pacs{05.50.+q} % Lattice theory and statistics

\maketitle

\section{Introduction}
\label{sec:intro}

The quest for better analytical and numerical methods in the
theoretical evaluation of measurable physical quantities, like
critical exponents and amplitude ratios, is one of the lasting tasks
of statistical field theory.

In recent years substantial progress in this field has been made by
the introduction of a method based on the strong coupling lattice
expansion of improved Hamiltonians \cite{CPRV-impr-O1}
(for a review, cf.\ Ref.\ \cite{PV-rept}).  The essential
feature of this method is the possibility of removing all leading
nonscaling contributions to physical quantities, in the neighborhood
of criticality, by a specific choice of a parameter in the lattice
Hamiltonian (critical coupling). The convergence of analytical and/or
numerical evaluations is therefore impressively faster than in any
other variant form of the models under inspection belonging to the
same universality class.

The main limitation of this method lays in the absence of an efficient
analytical technique for the determination of the critical parameter.
Conceptual reasons for this limitation may be found in the
impossibility of an $\epsilon$ expansion for the value of the critical
parameter, both in the $4-\epsilon$ and in the $2+\epsilon$ expansion
schemes.  In practice one must resort to an extrapolation from
numerical Monte Carlo finite-size evaluations of some physical
quantity, typically the Binder cumulant \cite{CPRV-impr-O1,PV-rept}.

Another well-known analytical approach to the study of critical
lattice models is the $1/N$ expansion, which applies in particular to
the physically very important class of $O(N)$ spin models in three
dimensions.  However it is known that, in exactly three dimensions and
for nearest-neighbor interactions, the critical parameter in the
large-$N$ limit is a negative number \cite{ZJ}.  Since the critical
parameter controls the large-field behavior of the interaction
potential, a negative value would naively imply an unbounded
Hamiltonian.  In practice this would prevent a Monte Carlo simulation
of the system, thus seriously jeopardizing the conceptual meaning of
the whole approach.

Nevertheless we still seriously believe the $1/N$ expansion to be,
from a theoretical point of view, probably the most relevant expansion
scheme that can be applied to any quantum and statistical field
theory, in that there is no known obstruction to summability of the
series expansion in powers of $1/N$ for the values of physical
quantities.

Therefore we decided to explore the conceptual and numerical
consequences of performing a systematic $1/N$ expansion of the
critical parameter, for the class of three-dimensional $O(N)$ spin
models, in order to check the actual relevance of the drawbacks that
we mentioned above.

As a consequence of our analysis we found that the critical parameter
can be formally computed within the expansion with no limitation related to the
sign of its large-$N$ value and for space dimensionalities in the
interval $ 2 < d < 4$. In particular we found that the sign of the
first $1/N$ correction is positive, and one may then expect to find a
value $N_c$ such that the parameter itself vanishes.  It is however
still unclear that the predictions of the $1/N$ expansion may be
extended to the region $N < N_c$.

As a byproduct of our analysis we obtained new more efficient
expressions for the asymptotic expansions of many important functions
entering our calculations. We presented these results with some
details because they might be relevant to other computations of
critical and subcritical quantities.

In Sec.\ \ref{sec:eff-H} we introduce the lattice $O(N)$ models and
their $1/N$ expansion. In Secs.\ \ref{sec:gap-eq}, \ref{sec:eff-prop},
and \ref{sec:renorm-g} we discuss on general grounds the relevant
asymptotic epansions (gap equation, effective propagator and
renomalized coupling). In Secs.\ \ref{sec:int-gap-eq} and
\ref{sec:int-eff-prop} we specialize these expansions to the case of
the standard nearest-neighbor interaction, with the help some useful
integral representations. A few numerical results are presented in
Sec.\ \ref{sec:numeric}.  In Sec.\ \ref{sec:improved} we compute the
$1/N$ correction to the improvement parameter and finally in Sec.\ 
\ref{sec:conclusions} we discuss the meaning and relevance of our
results.

\section{The effective Hamiltonian and the graph expansion}
\label{sec:eff-H}

Our starting point will be the usual $N$-component $\phi^4$ lattice
Hamiltonian in $d$ dimensions:
\begin{equation}
H = \sum_x \Biggl[{1 \over 2} {\textstyle \sum_\mu \nabla_\mu
{\bm\phi}(x) \cdot \nabla_\mu {\bm\phi}(x)} + {1 \over 2} \mu_0^2 
{\bm\phi}^2(x) + {1 \over 4!} g_0 \bigl({\bm\phi}^2(x)\bigr)^2 \Biggr],
\label{eq:H}
\end{equation}
where $ \nabla_\mu {\bm\phi}(x)$ is some (local) form of the lattice
gradient; in the standard nearest-neighbor formulation 
$\nabla_\mu {\bm\phi}(x) = {\bm\phi}(x{+}\mu)- {\bm\phi}(x)$.

Following Ref.\ \cite{CPRV-ON} we define the rescaled couplings
\[
\beta = - {6 \mu_0^2 \over g_0 N},\qquad
\gamma = {3 \over g_0 N}
\]
and introduce an auxiliary field $\alpha$ in order to eliminate the
quartic term in the Hamiltonian.

After a trivial Gaussian integration the resulting effective Hamiltonian is
\begin{equation}
H_{\rm eff} = {N \over 2} [\Tr\Ln \beta (-\nabla_\mu \nabla_\mu
+ i \alpha)- i \beta \alpha + \gamma \alpha^2 ].
\label{eq:Heff}
\end{equation}

In the limit $\gamma \to 0$ this Hamiltonian reduces to the usual
effective large-$N$ expression for the nonlinear $\sigma$ model.  In
the nearest-neighbor formulation, it is also known as the $O(N)$
Heisenberg model; its large-$N$ limit was investigated in Ref.\ 
\cite{MR}.

The saddle-point condition on the effective Hamiltonian leads to the
so-called gap equation 
\begin{equation}
\beta + 2 \gamma m_0^2 = \int_{-\pi}^\pi {d^d p \over (2 \pi)^d}
        { 1 \over \bar p^2 + m_0^2},
\label{eq:saddle}
\end{equation}
where $\bar p^2$ is the Fourier transform of the lattice Laplacian
operator $-\nabla_\mu \nabla_\mu$, which in the nearest-neighbor case
takes the form $\hat p^2 = 2 \sum_\mu (1 - \cos p_\mu)$.

The gap equation allows for the elimination of $\beta$ in favor of the
new parameter $m_0$  (large-$N$ inverse correlation length) in the
Feynman graph expansion.

In the large-$N$ limit criticality corresponds to the vanishing of
$m_0^2$, and the criticality condition may then take the form
\[
\beta_c =  \int_{-\pi}^\pi {d^d p \over (2 \pi)^d} {1 \over \bar p^2},
\]
corresponding to a finite value of $\beta_c$ for all $d>2$.

The graph expansion for this model, in the formulation based on the
effective Hamiltonian, requires defining the (bare) propagator
$\Delta$ for the effective field $\alpha$; by standard manipulations
one obtains
\begin{equation}
D (k, m_0, \gamma) \equiv \Delta^{-1}(k, m_0, \gamma) = 
{1 \over 2} \int_{-\pi}^\pi {d^d p \over (2 \pi)^d}
        {1 \over \bar p^2 +m_0^2}{1 \over ({\overline {p+k}})^2 +m_0^2}
        + \gamma.
\label{eq:Delta}
\end{equation}
The above approach is quite general, and it leads to a systematic
$1/N$ expansion of the correlation functions and of the physical
quantities for arbitrary values of $m_0$ and $\gamma$.

However we want to focus our attention on the critical domain, and in
particular we want to evaluate the coupling $\gamma^*$ such that the
first nontrivial corrections to scaling turn out to vanish in the
computation of physical quantities in the scaling region.

To this purpose it is convenient to parametrize the cutoff dependence
of correlation functions and renormalized couplings in terms of the
lattice spacing $a$, which can be made to appear explicitly in
calculations by a rescaling of the coupling and momentum dependence.

Around criticality the dependence on $a$ is not analytic, and as a
consequence we need asymptotic expansions in order to identify the
scaling and leading nonscaling contributions to any computable
quantity.

The basic technique for asymptotic expansions in powers of $m_0 a$ is
described in Ref.\ \cite{CR}; here we shall discuss its applications
to the cases of interest for the present paper.

We only recall that, in order to regularize the generic lattice integral
\[
I(k; m_0 a) \equiv \int_{-\pi}^\pi {d^d p \over (2 \pi)^d} F(k; m_0 a, p),
\]
where $k$ is any collection of external momenta, we can make use of the 
formal identity
\[
I(k; m_0 a) = I_{\rm lat}(k; m_0 a)+I_{\rm con}(k; m_0 a),
\]
where 
\[
I_{\rm lat}(k; m_0 a) =
\int_{-\pi}^\pi {d^d p \over (2 \pi)^d} F(k; m_0 a, p)
- \int_{-\infty}^{+\infty} a^d{d^d p \over (2 \pi)^d} 
    T^{\rm(IR)} F(k; m_0 a, pa) 
\]
and
\[
I_{\rm con}(k; m_0 a)= a^d \int_{-\infty}^{+\infty}
 {d^d p \over (2 \pi)^d} [F(k; m_0 a, pa)- T^{\rm(UV)} F(k; m_0 a, pa)].
\]

The $T \equiv T^{\rm(IR)}+T^{\rm(UV)}$ operation amounts to a Taylor
series expansion of the integrand in powers of $m_0 a$, $T^{\rm(IR)}$
and $T^{\rm(UV)}$ corresponding respectively to the IR and UV singular
terms in the expansion.

It is possible to prove that the expansions of $I_{\rm lat}$ (the
``lattice contribution'') and $I_{\rm con}$ (the ``continuum
contribution'') are individually and fully regular; the nonanaliticity
of the expansion is factored out in the $a^d$ term multiplying the
continuum contribution.

\section{Asymptotic expansion of the gap equation}
\label{sec:gap-eq}

For the purposes of the present paper and in order to show an explicit
example of the asymptotic expansion procedure let us consider the gap
equation in the large-$N$ limit.

Let us assume a generic lattice Laplacian such that 
$\bar p^2 \approx p^2 +c\, p^4+O(p^6)$, where 
$p^{2n}=\sum_\mu p_\mu^{2n}$: in the nearest-neighbor version 
$c = -{1 \over 12}$.

From the previously derived results we obtain the following relationship:
\[
\beta_c -\beta = m_0^2 a^2 \left[\int_{-\pi}^\pi {d^d p \over (2 \pi)^d}
{1 \over \bar p^2 (\bar p^2+ m_0^2 a^2)}+ 2 \gamma \right].
\]

Let us now define 
\[
G(p, m_0 a) \equiv   {m_0^2 a^2 \over \bar p^2 (\bar p^2+ m_0^2 a^2)}
\]
and perform the relevant expansions up to the first few nontrivial terms:
\begin{eqnarray*}
T\, G(p, m_0 a) &\approx& {m_0^2 a^2 \over (\bar p^2)^2} 
- {m_0^4 a^4 \over (\bar p^2)^3}+ O\bigl(m_0^6 a^6\bigr), \\
G(p a, m_0 a) &\approx& {1 \over a^2}
\left[{1 \over p^2}-{1 \over p^2+m_0^2}\right]
- c\, p^4\left[{1 \over (p^2)^2} - {1 \over (p^2+m_0^2)^2}\right]
+ O\bigl(a^2\bigr), \\
T\,G(p a, m_0 a) &\approx& {m_0^2 \over a^2}
\left[{1 \over (p^2)^2}- {m_0^2 \over (p^2)^3}\right]
- m_0^2\, c\, p^4 \left[{2 \over (p^2)^3}-{3 m_0^2 \over (p^2)^4}\right]
+ O\bigl(m_0^6, a^2 \bigr).
\end{eqnarray*}

By grouping together the IR singular terms we therefore obtain the
lattice contributions:
\[
2 m_0^2 a^2 (\gamma - \gamma_0) -m_0^4 a^4 \delta_0 
+ O \bigl(m_0^6 a^6\bigr),
\]
where we defined the following numerical constants:
\begin{eqnarray}
\gamma_0 &\equiv& {1 \over 2}
\left[- \int_{-\pi}^\pi {d^d p \over (2 \pi)^d} {1 \over (\bar p^2)^2}
      + \int_{-\infty}^{+\infty} {d^d p \over (2 \pi)^d}{1 \over (p^2)^2} 
\right], \nonumber \\
\delta_0 &\equiv& 
  \int_{-\pi}^\pi {d^d p \over (2 \pi)^d} {1 \over (\bar p^2)^3}
- \int_{-\infty}^{+\infty} {d^d p \over (2 \pi)^d} 
      \left({1 \over (p^2)^3}-3 c\,{p^4 \over (p^2)^4}\right).
\label{eq:gamma0,delta0}
\end{eqnarray}

On the other side, by grouping together the UV singular terms, we
obtain the following continuum contributions:
\[
\int_{-\infty}^{+\infty} {d^d p \over (2 \pi)^d} \left(a^{d-2}
    \left[-{1 \over p^2+m_0^2}+{1 \over p^2}\right]
    + a^d c\,p^4 \left[{1 \over (p^2+m_0^2)^2}
        - {1 \over (p^2)^2}-2{m_0^2 \over (p^2)^3}\right]
    + O(a^{d+2})\right).
\]

We can perform the continuum integrals by standard dimensional
regularization techniques. We then finally find 
\begin{equation}
\beta_c -\beta \approx -b_0\, (m_0 a)^{d-2}+ 2 (\gamma -\gamma_0) m_0^2 a^2
- {3 \over 2}c\,b_0\, m_0^d a^d -\delta_0\, m_0^4 a^4 
+ O\bigl( (m_0 a)^{d+2}\bigr),
\label{eq:beta-cont}
\end{equation}
where 
\[
b_0 \equiv {\Gamma(1-{d \over 2}) \over (4 \pi)^{d \over 2}}.
\]

From the asymptotic expansion of the gap equation we immediately learn
the following lesson: it is possible to choose for the quartic
coupling $\gamma$ the special value $\gamma_0$ such that the first
nonscaling contribution to the large-$N$ saddle-point condition
vanishes.

By this procedure we have identified the large-$N$ critical coupling
for these versions of the model. In fact any change in the form of the
local interaction, reflecting itself in the detailed $p$ dependence of
the bare lattice massless propagator without changing its singular
part, leads to a finite change in the numerical value of $\gamma_0$
without affecting its formal representation.

It is especially interesting to consider the class of models
characterized by next-to-nearest neighbor interactions, and whose
propagator of the fundamental excitations is obtained in the form
\[
\bar p^2 = \sum_{\mu=1}^d {\textstyle\bigl(6 c+{5 \over 2}\bigr)
- 8 \bigl(c+{1\over 3}\bigr)\cos p_\mu
+ 2 \bigl(c+{1 \over 12}\bigr)\cos 2 p_\mu} =
\hat p^2+\bigl(c+{\textstyle{1\over12}}\bigr)\hat p^4,
\]
where
\[
\hat p^n \equiv \sum_{\mu=1}^d \hat p_\mu^n, \qquad
\hat p_\mu \equiv 2 \sin {p_\mu \over 2}.
\]
Note that $c=-{1\over12}$ corresponds to the standard nearest-neighbor
interaction, while $c=0$ corresponds to the $O(a^2)$ Symanzik
tree-improved version of $O(N)$ models \cite{Symanzik}.

\begin{figure}[tb]
\begin{center}
\leavevmode
\epsfig{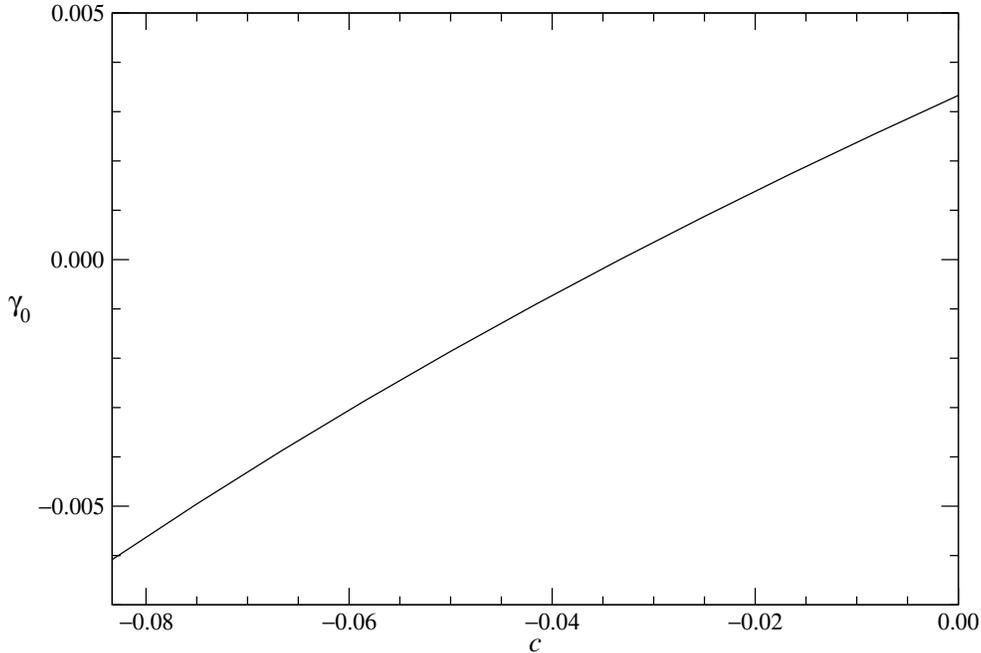}
\vskip-7mm
\end{center}
\caption{$\gamma_0$ vs.\ $c$.
}
\label{fig:gamma0}
\end{figure}

\begin{figure}[tb]
\begin{center}
\leavevmode
\epsfig{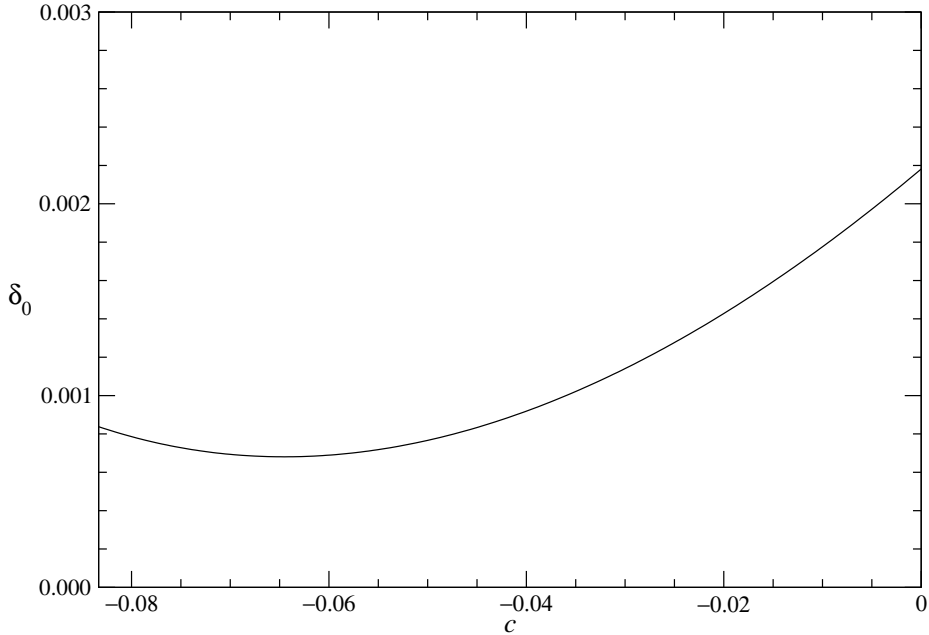}
\vskip-7mm
\end{center}
\caption{$\delta_0$ vs.\ $c$.
}
\label{fig:delta0}
\end{figure}

In three dimensions we have numerically explored the range 
$-{1\over12}\le c\le0$: our results are presented in Figs.\ \ref{fig:gamma0}
and \ref{fig:delta0}.

Let us notice in particular that the choice $c = -0.03332110...$
corresponds to vanishing $\gamma_0$, and it is therefore, at least in
the large-$N$ limit, an alternative version of a spin model where the
leading corrections to scaling are automatically made to vanish.  In
turn when $c=0$ we obtain $\gamma_0 = 0.003328210...$, which implies a
small but nonvanishing leading scaling violation, and 
$\delta_0 = 0.002181406...$ .  Compared to the standard
$c=-{1\over12}$ case this version has however the advantage of being
numerically testable also by Monte Carlo methods, since $\gamma_0>0$.

\section{Asymptotic expansion of the effective propagator}
\label{sec:eff-prop}

In order to perform an asymptotic expansion of the $O(1/N)$
contributions to physical quantities we must compute one-loop graphs
involving the effective propagator $\Delta (k, m_0, \gamma)$.

We therefore need to evaluate the asymptotic expansion of $\Delta$ or,
more conveniently, of $D(k, m_0, \gamma)$.  It is easy to recognize,
from the definition of $D$ and from the general rule of the asymptotic
expansion, that in general we may express the result in the form
\begin{equation}
D(k, m_0 a, \gamma) = \sum_{n=0}^\infty 
\bigl [A_n (k) (m_0 a)^{2n} + B_n (k) (m_0 a)^{2n+d-2} \bigr],
\label{eq:D-expansion}
\end{equation}
where $A_n(k)$ have the form of lattice contributions (and only $A_0$
depends on $\gamma$), while $B_n(k)$ are continuum contributions that
can be analytically computed, e.g., in dimensional regularization.

We shall not repeat the derivation (some details can be found in Ref.\
\cite{CR}), but only quote the relevant results:
\begin{eqnarray}
A_0(k,\gamma) &=& {1 \over 2} \int_{-\pi}^\pi {d^d p \over (2 \pi)^d}
{1 \over \bar p^2}{1 \over ({\overline {p+k}})^2} + \gamma, \nonumber \\
A_1(k) &=& -\int_{-\pi}^\pi {d^d p \over (2 \pi)^d} {1 \over (\bar p^2)^2}
{1 \over ({\overline {p+k}})^2}+\int_{-\infty}^{+\infty}
    {d^d p \over (2 \pi)^d}{1 \over (p^2)^2}{1 \over \bar k^2}, \nonumber \\
A_2(k) &=& \int_{-\pi}^\pi {d^d p \over (2 \pi)^d} {1 \over (\bar p^2)^3}
{1 \over ({\overline {p+k}})^2}+{1 \over 2}{1 \over (\bar p^2)^2}
    {1 \over (({\overline {p+k}})^2)^2}
- \int_{-\infty}^{+\infty}{d^d p \over (2 \pi)^d}{1 \over (p^2)^3}
{1 \over \bar k^2} \nonumber \\
&-& 3 c{p^4 \over (p^2)^4}{1 \over \bar k^2}
+ {1 \over (p^2)^2}C_0(k)+{1 \over 2}\left({1 \over (p^2)^2}
+ {1 \over ((p+k)^2)^2}\right){1 \over (\bar k^2)^2}; \nonumber \\
B_0(k) &=& {b_0 \over \bar k^2}, \qquad B_1(k) 
= -b_0\left({1 \over (\bar k^2)^2} + C_0(k)
- {3 \over 2}{c \over \bar k^2}\right),
\label{eq:Bn}
\end{eqnarray}
where we defined
\[
C(k, m_0) \equiv {1 \over 2 d}{\textstyle \sum_\mu} 
{\partial^2 \over \partial k_\mu^2}{1 \over \bar k^2+ m_0^2}, \qquad
C_0(k) \equiv C(k,0) \equiv {1 \over 2 d}{\textstyle \sum_\mu}
{\partial^2 \over \partial k_\mu^2}{1 \over \bar k^2}.
\]

Trivial manipulations allow to express $A_n(k)$ as  pure lattice integrals:
\begin{eqnarray}
A_1(k) &=& \int_{-\pi}^\pi {d^d p \over (2 \pi)^d}{1 \over (\bar p^2)^2}
\left[{1 \over \bar k^2}-{1 \over ({\overline {p+k}})^2}\right] 
+ 2 {\gamma_0 \over \bar k^2}, \nonumber \\
A_2(k) &=& \int_{-\pi}^\pi {d^d p \over (2 \pi)^d} 
\left[{1 \over (\bar p^2)^3}{1 \over ({\overline {p+k}})^2}
    - {1 \over (\bar p^2)^3}{1 \over \bar k^2}
    - {1 \over (\bar p^2)^2}C_0(k)\right]
+ {\delta_0 \over \bar k^2} -2 \gamma_0\, C_0(k)
\label{eq:An-latt} \\
&+& \int_{-\pi}^\pi {d^d p \over (2 \pi)^d} 
\left[{1 \over 2}{1 \over (\bar p^2)^2}
    {1 \over (({\overline {p+k}})^2)^2}
    - {1 \over 2}{1 \over (\bar p^2)^2}{1 \over (\bar k^2)^2}
    - {1 \over 2}{1 \over (({\overline {p+k}})^2)^2}
    {1 \over (\bar k^2)^2}\right] - 2{\gamma_0 \over (\bar k^2)^2}. 
\nonumber
\end{eqnarray}

From our general considerations it should be by now clear that we
shall also need a different expansion of $D$, homogeneous in powers of
$m$ and $k$.

Without delving into the details, we find that
\[
D(k a, m_0 a, \gamma) \approx
{1 \over 2} a^{d-4} \int_{-\infty}^{+\infty}{d^d p \over (2 \pi)^d}
{1 \over p^2 +m_0^2} {1 \over (p+k)^2+m_0^2} +(\gamma-\gamma_0) + O(a^{d-2}).
\]

Notice that the integral can be analytically computed in all
dimensions, and the final result is
\begin{equation}
D(k a, m_0 a, \gamma) \approx 
d_0 \left({1 \over 4}k^2 a^2+ m_0^2 a^2\right)^{{d \over 2}-2}
{}_2F_1\!\left(2-{d \over 2},{1 \over 2}, {3 \over 2}, {1 \over \xi^2}\right)
+ \gamma-\gamma_0 \equiv D_c,
\label{eq:Dc}
\end{equation}
where $d_0 = {1 \over 2}(1-{d \over 2})b_0$, ${}_2F_1$ is the Gauss
hypergeometric function and $\xi = \sqrt{1 + 4 m_0^2/k^2}$.

Let us in general denote by the label ``$c$'' the quantities occurring in
the leading order in the homogeneous (continuum) expansion of $D$, and
in particular $\gamma_c \equiv \gamma -\gamma_0$.

Following Ref.\ \cite{DRG}, we can exploit identities between
hypergeometric functions to recast the above result into the form
\begin{equation}
D_c = a_0\, \xi^{d-3} (k a)^{d-4} + \gamma_c + 
{b_0 \over k^2 a^2 \xi^2} \,
{}_2F_1\!\left({d-1 \over 2},1,{d \over 2},
    1-{1 \over \xi^2}\right)(m_0 a)^{d-2},
\label{eq:Dc1}
\end{equation}
where
\[
a_0 \equiv {1 \over 2}{\Gamma({d \over 2}-1)^2 \Gamma(2-{d \over 2})
\over (4 \pi)^{d \over 2} \Gamma(d-2)}.
\]
For $d=3$, Eq.\ (\ref{eq:Dc1}) reduces to
\[
D_c = {1\over16\,ak} + \gamma_c - {1\over8\pi ak}\,\arccot{k\over2m_0}.
\]

Eq.\ (\ref{eq:Dc1}) is especially appropriate for the asymptotic
expansion of $D_c$, because all the nonanalytic dependence on $m_0 a$
is explicitly factored out in the last term.  In particular (after a
rescaling $ka \to k$) we obtain the following behaviors:
\[
A_{0c} = a_0 (k^2)^{{d \over 2}-2}+\gamma_c,\qquad  
A_{nc} = {1 \over n!}{4^n \Gamma({d-1 \over 2}) \over
    \Gamma({d-1 \over 2}-n)} a_0 (k^2)^{{d \over 2}-2-n}.
\]

Finally notice also that the zero-momentum value of $D$ is related to
the derivative of the gap equation with respect to the mass, and we
can obtain the relationship
\begin{equation}
D_0 \equiv D(0, m_0 a, \gamma) \approx d_0\, (m_0 a)^{d-4} + 
\gamma_c - {3 d \over 8} c\, b_0\,(m_0 a)^{d-2} - \delta_0\, m_0^2 a^2 ,
\label{eq:D0}
\end{equation}
and as a consequence
\[
D_{0c} \equiv d_0 (m_0 a)^{d-4} +\gamma_c.
\]

\section{Asymptotic expansion of the renormalized coupling}
\label{sec:renorm-g}

We now recall from the literature the expression of the $O(1/N)$
contribution to the (unrenormalized) self-energy of the fundamental
quanta:
\[
\Sigma_1(p,m_0) = \Sigma_{1a}(p,m_0)+\Sigma_{1b}(p,m_0),
\]
where
\begin{eqnarray*}
\Sigma_{1a}(p,m_0) &=& \int_{-\pi}^\pi {d^d k \over (2 \pi)^d}
{\Delta(k,m_0,\gamma) \over ({\overline {p+k}})^2+m_0^2}, \\
\Sigma_{1b}(m_0) &=& {1 \over 2} \Delta(0,m_0,\gamma)
\int_{-\pi}^\pi {d^d k \over (2 \pi)^d}\Delta(k,m_0,\gamma)
{\partial \over \partial m_0^2}\Delta^{-1}(k,m_0,\gamma).
\end{eqnarray*}

Again we might perform an asymptotic expansion of this expression, on
the lines traced in Ref.\ \cite{CR}, recovering in the scaling limit
the (unrenormalized) continuum contribution, and in principle we might
evaluate the first nonleading contribution.

However for our purposes it is much more convenient to work directly
with quantities chosen in such a way that all renormalization effects
are automatically removed, i.e., quantities whose scaling limit is a
finite, renormalization-group invariant, amplitude.

The simplest such object is the so-called ``renormalized coupling''
$g_r$, whose continuum (scaling) value $g_r^*$ has been computed to
$O(1/N)$ in Ref.\ \cite{CPRV-ON}.

Actually the formal expression derived in Ref.\ \cite{CPRV-ON} is
correct also for the lattice versions of the model, when continuum
propagators are replaced by their lattice counterparts. Parametrizing
the result in terms of the renormalized mass $m$ we therefore obtain
\begin{equation}
g_r(m,\gamma) = m^{d-4} \Delta(0,m,\gamma)
\left[1+{1 \over N} g_r^{(1)}(m, \gamma)
    + O\left({1 \over N^2}\right)\right],
\label{eq:Gr-largeN}
\end{equation}
where $g_r^{(1)}$ is given by
\begin{eqnarray}
g_r^{(1)}(m, \gamma) &=& \Delta(0.m,\gamma)
{\partial \Delta^{-1}(0,m,\gamma) \over \partial m^2}
\left(\Sigma_{1a}(0,m)+\Sigma_{1b}(m)
- m^2{\partial \Sigma_{1a} \over \partial p^2}\Bigg\vert_0\right)
\nonumber \\
&+& \left(2 {\partial \Sigma_{1a}(0,m) \over \partial m^2} +
{\partial \Sigma_{1b}(m) \over \partial m^2}
- 2 {\partial \Sigma_{1a} \over \partial p^2}\Bigg\vert_0 
- 2 \Delta^{-1}(0,m,\gamma) T(m)\right),
\label{eq:g1r}
\end{eqnarray}
and  we  defined
\[
T(m_0)= \int_{-\pi}^\pi {d^d k \over (2 \pi)^d}
\left({\Delta(k,m_0,\gamma) \over \bar k^2+m_0^2}\right)^2.
\]

It is now a matter of trivial algebraic manipulations to show that the
expression for $g_r^{(1)}$ can be cast into the form
\begin{eqnarray}
g_r^{(1)} &=& {1 \over D_0}\int \limits_{-\pi}^\pi 
{d^d k \over (2 \pi)^d} \Biggl\{{1 \over D}\left({1 \over 2}
{\partial \over \partial m^2}
\left[{\partial D \over \partial m^2}+ {2 D_0 \over \bar k^2+m^2}\right]
- { D_0 \over (\bar k^2+m^2)^2} - \left[2 D_0+ m^2{\partial D_0
   \over \partial m^2}\right] C\right) \nonumber \\
&&\qquad\qquad\qquad-\, {1 \over 2\,D^2}\left[{\partial D \over \partial m^2}
+ { 2 D_0 \over \bar k^2+m^2}\right]^2\Biggr\}.
\label{eq:g1r-full}
\end{eqnarray}

In the asymptotic expansion of $g_r^{(1)}$ we may again identify a
continuum and a lattice contribution.

Renormalization-group theory insures us about the expected properties
of these contributions. In particular the leading continuum term
should be finite and should therefore require no UV counterterms. Its
evaluation will however allow us to identify the IR counterterms
needed in order to regularize the lattice term.

The lattice contribution in turn should not affect the leading
(scaling) order, since $g_r^*$ is an invariant amplitude which should
not depend on the detailed form of the Hamiltonian.  In turn, trivial
power counting in $a$ shows us that the first correction to scaling is
generated by the leading lattice contribution, which is
$O\bigl((ma)^{4-d}\bigr)$.

We can prove the following identities, corresponding to similar
results of  Ref.\ \cite{CPRV-ON}:
\begin{eqnarray*}
{\partial D_c \over \partial m^2} &=&
{2 \over k^2+4 m^2}\left[(d-3)D_c -D_{0c}+(4-d)\gamma_c \right], \\
{\partial D_{0c} \over \partial m^2} &=& 
\left({d \over 2}-2\right){D_{0c}-\gamma_c \over m^2}, \\
C_c &=& {1 \over (k^2+m^2)^2}\left({4 \over d}{k^2 \over k^2+m^2}-1\right).
\end{eqnarray*}

As a consequence we obtain in leading order the following continuum
contribution, depending on the single (dimensionless) variable 
$x \equiv \gamma_c (ma)^{4-d}$:
\begin{eqnarray}
g_{r,\rm con}^{(1)}(x) &\approx& 
{1 \over D_{0c}}\int_{-\infty}^{+\infty} {d^dk \over (2 \pi)^d} \nonumber \\
\times\,\Biggl(\!&-&\!2\left[{d-3 \over k^2+4m^2}
+ {\gamma_c \over D_c}{4-d \over k^2+4 m^2}
+ {D_{0c} \over D_c}{3 m^2 \over (k^2+m^2)(k^2+4 m^2)} \right]^2 
+ 2{(d-3)(d-5) \over (k^2+4 m^2)^2} \nonumber \\
&+& {\gamma_c (4-d) \over D_c}
  \left[{2(d-5) \over (k^2+4 m^2)^2}
    + {3 \over 2}{1 \over(k^2+4 m^2)(k^2+m^2)}
    + {1 \over 2}\left(1 - {4 \over d}\right){1 \over( k^2+m^2)^2} \right]
\nonumber \\
&+& {2\gamma_c\over D_c}{ (4-d) m^2 \over d(k^2+m^2)^3} \nonumber \\
&+& {D_{0c}\over D_c}{m^2 \over k^2+m^2} 
  \left[{6(d-5) \over( k^2+4 m^2)^2}
    + {3 \over 2}{d-8 \over(k^2+4 m^2)( k^2+m^2)}
    + {2 \over (k^2+m^2)^2} \right]\Biggr),
\label{eq:g1r,con}
\end{eqnarray}
which can be shown to correspond exactly to the result presented in
Ref.\ \cite{CPRV-ON}. 

In particular the fixed-point value is obtained by setting $x=0$,
corresponding to $\gamma_c =0$, that is the condition for the removal
of the leading nonscaling behavior in the large-$N$ limit.

From the above result we may immediately read off the structure of
counterterms, just by taking the power series expansion in powers of
$m^2$, and in particular by replacing $D_c$ with $A_{0c}$ and $D_{0c}$
with $d_0 (ma)^{d-4}$.  The resulting (singular) expression is
\[
{(ma)^{4-d} \over d_0}\int\limits_{-\infty}^{+\infty} {d^dk \over (2 \pi)^d} 
\left(-2 \left[{d-3 \over k^2}+{\gamma_c \over A_{0c}}{4-d \over k^2}\right]^2
    + 2{(d-3)(d-5) \over (k^2)^2}
    + 2{\gamma_c \over A_{0c}}
      \left(d-4-{1 \over d}\right){4-d \over (k^2)^2}\right).
\]

Let us now compute the lattice contribution.
By applying the already defined asymptotic expansions we obtain:
\[
{\partial D \over \partial m^2}+{2 D_0 \over \bar k^2+m^2} \approx
A_1+ 2 {\gamma_c \over \bar k^2} 
- b_0 \left({d \over 2}C_0 + {1 \over (\bar k^2)^2} \right)m^{d-2}
+ 2\left(A_2-{\gamma_c \over (\bar k^2)^2}-{\delta_0 \over \bar k^2}\right)m^2
+ O(m^d)
\]
and
\begin{eqnarray*}
&&{1 \over 2}{\partial \over \partial m^2}
\left[{\partial D \over \partial m^2}+{2 D_0 \over \bar k^2+m^2}\right]
- { D_0  \over (\bar k^2+m^2)^2}
- \left[2\,D_0+ m^2{\partial D_0 \over \partial m^2}\right] C \\
&\approx&
A_2-{\delta_0 \over \bar k^2}
- 2 \gamma_c \left(C_0+{1 \over (\bar k^2)^2}\right)+ O(m^{d-2}).
\end{eqnarray*}
Notice that the most singular contributions, $O(m^{d-4})$, have been
removed in the above combinations.

Let us now find out the leading IR singularities of the terms appearing
in the integral representing $g_{r,\rm lat}^{(1)}$. We must only notice
that, for all $A_n(k)$, the singular behavior when $k \to 0$
is determined by the corresponding behavior of $A_{nc}(k)$, with
corrections whose singularity is depressed by a factor $k^2$. As a
consequence we obtain in the IR limit
\begin{eqnarray}
{1 \over A_0}\left(A_1+2 {\gamma_c \over k^2}\right) &\to&
{2 \over k^2}\left[d-3  +(4-d){\gamma_c \over A_{0c}}\right],
\label{eq:An-IR} \\
{1 \over A_0}\left(A_2-{\delta_0 \over \bar k^2}- 2 \gamma_c (C_0+{1
\over (\bar k^2)^2})\right) &\to& {2 \over (k^2)^2}
\left[(d-3)(d-5) +
    \left(d-4-{1 \over d}\right)(4-d){\gamma_c \over  A_{0c}}\right].
\nonumber
\end{eqnarray}

These singularities are perfectly matched by the terms coming from the
$T^{\rm(IR)}$ expansion of the continuum contribution, as expected.
As a consequence we are able to write down an exact, finite
representation of the leading lattice contribution to $g_r^{(1)}$,
taking the form
\begin{equation}
g_{r,\rm lat}^{(1)}(m, \gamma_c) \equiv 
{(m a)^{4-d} \over d_0} \delta g^{(1)}(\gamma_c),
\label{eq:g1r,lat}
\end{equation}
where
\begin{eqnarray}
\delta g^{(1)}(\gamma_c) &\equiv&
\int_{-\pi}^\pi {d^d k \over (2 \pi)^d} 
\Biggl({1 \over A_0}\left(A_2-{\delta_0 \over \bar k^2}
    - 2 \gamma_c \left(C_0+{1 \over (\bar k^2)^2}\right)\right)
- {1 \over 2}\left[{1 \over A_0}
    \left(A_1+2 {\gamma_c \over \bar k^2}\right)\right]^2
\Biggr) \nonumber \\
&-& \int \limits_{-\infty}^{+\infty} {d^d k \over (2 \pi)^d} 
{2 \over (k^2)^2} \Biggl(\left[ (d-3)(d-5)
    + \left(d-4-{1 \over d}\right)(4-d){\gamma_c \over  A_{0c}}\right]
\nonumber \\ &&\qquad\qquad\qquad-\, 
\left[d-3  +(4-d){\gamma_c \over A_{0c}}\right]^2 \Biggr).
\label{eq:delta-g1r}
\end{eqnarray}

It is worth observing that $g_{r,\rm con}^{(1)}(x)$ showed a
nonanalyticity whose leading dependence was proportional to $x \ln x$.
Correspondingly $\delta g^{(1)}(\gamma_c)$ has a nonanalytic $\gamma_c
\ln \gamma_c$ dependence. The coefficients of these singularities
match properly in order to reproduce an overall dependence
proportional to $\gamma_c (ma)^{4-d} \ln ma$, as expected from general
renormalization group arguments because of the anomalous dimension of
the leading irrelevant operator.

\section{Integral representation for the gap equation}
\label{sec:int-gap-eq}

For the purpose of actual numerical calculations one must find an
efficient way of performing lattice momentum integrals. In practice
this may be obtained by resorting to parametric (Feynman and
Schwinger) representations of the lattice propagators. These
representations are especially useful in the case corresponding to the
standard nearest-neighbor Hamiltonian.

Let us first consider the integral appearing in the gap equation
\[
\beta + 2 \gamma m_0^2 =  \int_{-\pi}^\pi {d^d p \over (2 \pi)^d}
{ 1 \over \hat p^2 + m_0^2} \equiv \Xi (m_0^2).
\]

Introducing Schwinger's proper time representation we obtain
\begin{equation}
\Xi (m_0^2) = \int_0^\infty d\alpha\, e^{-\alpha m_0^2} 
\int_{-\pi}^\pi {d^d p \over (2 \pi)^d} e^{-2 \alpha \sum_\mu (1-\cos p_\mu)} 
= \int_0^\infty d\alpha\, e^{-\alpha m_0^2} 
\left[e^{-2 \alpha} I_0(2 \alpha)\right]^d, 
\label{eq:Xi-integral-repr}
\end{equation}
where $I_0$ is the standard modified Bessel function, admitting for
large values of its argument the following asymptotic expansion:
\[
e^{-2 \alpha}I_0(2\alpha) \approx 
{1 \over(4 \pi \alpha)^{1 \over 2}}\sum_{n=0}^\infty  
{(-1)^n\Gamma(n+{1 \over 2}) \over 
 n!\,\Gamma({1 \over 2}-n)}{1 \over (4\alpha)^n}.
\]
By a proper change of variables the numerical evaluation of the above
integral is now possible even in the small $m_0^2$ regime.

When $d=3$ it is also possible to obtain an analytical expression for
$\Xi(0)$ \cite{MR}, which was first derived in Ref.\ \cite{Watson}:
\begin{equation}
\beta_c = \Xi (0) = {\varkappa^2+1 \over \pi^2} K^2 (\varkappa),
\label{eq:betac}
\end{equation}
where $K$ is the complete elliptic integral of the first kind and
$\varkappa = (2-\sqrt{3})(\sqrt{3}-\sqrt{2})$, and the resulting
numerical value is $\beta_c = 0.252731009858663...$

We can compute $\Xi(m_0^2)$ as an asymptotic expansion around $m_0^2 =0$:
\begin{equation}
\Xi (m_0^2) =\sum_{n=0}^\infty a_n m_0^{2n}+b_n m_0^{2n+d-2}.
\label{eq:Xi-asympt}
\end{equation}
It is conceivable that when $d=3$ the coefficients $a_n$ of the
expansion can be computed analytically in terms of elliptic integrals,
but we contented ourselves with a numerical calculation of the
``lattice terms'' appearing in the analytic part of the expansion,
while it is always possible to obtain closed form expressions for the
coefficients $b_n$ of the nonanalytic part.

We can obtain explicit expressions for the coefficients in the
expansion (\ref{eq:Xi-asympt}) by subtracting a proper number of terms
of the asymptotic expansion of the Bessel function raised to the power
$d$.  Let us label the coefficients of this expansion according to the
equation
\begin{equation}
\Bigl[ e^{-2 \alpha}I_0(2\alpha) \Bigr]^d \approx 
{1 \over (4 \pi\,\alpha)^{d \over 2}} \sum_{n=0}^\infty c_n(d)\,\alpha^{-n};
\label{eq:I0-asympt}
\end{equation}
the coefficients can be computed recursively from the equations
\begin{eqnarray}
c_0(d) &=& 1, \nonumber \\
c_n(d) &=& {1\over n} \sum_{k=1}^n (kd - n + k)
{(-1)^k \, \Gamma({\textstyle{1\over2}}+k) \over
 4^k k! \, \Gamma({\textstyle{1\over2}}-k)} \, c_{n-k}.
\label{eq:c_n}
\end{eqnarray}

Let us add to the integrand in Eq.\ (\ref{eq:Xi-integral-repr}) the
formally vanishing term
\[
- {1 \over (4 \pi\,\alpha)^{d \over 2}} 
    \sum_{n=0}^\infty {(-\alpha\,m_0^2)^n \over n!} 
    \sum_{m=0}^{n-1} c_m(d)\, \alpha^{-m} 
+ {1 \over (4 \pi\,\alpha)^{d \over 2}} 
    \sum_{n=0}^\infty c_n(d)\,\alpha^{-n} 
    \sum_{m=n+1}^\infty {(-\alpha\,m_0^2)^m \over m!},
\]
where we have only interchanged the order of the summations in the two
contributions.  It is now possible to group the first contribution
with the original integrand and recognize that the resulting
combination defines an analytic function of $m_0^2$, since each
coefficient
\begin{equation}
a_n \equiv \int_0^\infty d \alpha\, {(-\alpha)^n \over n!} 
\Biggl\{\Bigl[e^{-2 \alpha}I_0(2\alpha) \Bigr]^d 
- {1 \over (4 \pi\,\alpha)^{d \over 2}} 
\sum_{m=0}^{n-1} c_m(d) \,\alpha^{-m} \Biggr\}
\label{eq:a_n}
\end{equation}
in the corresponding power series expansion is a finite, UV and IR
regulated integral for any $d$ in the range $2 < d < 4$.  In turn, the
integration of the second contribution may be represented, after
trivial resummations and rescalings, in the form
\[
{1 \over (4 \pi)^{d \over 2}} 
\sum_{n=0}^\infty c_n(d) \, k_n(d) \, (m_0^2)^{n+{d \over 2}-1},
\]
where
\[
k_n(d) \equiv \int_0^\infty dx\,x^{-n-{d \over 2}} 
\Biggl[e^{-x} - \sum_{m=0}^n {(-x)^m \over m!} \Biggr]
\]
are finite UV and IR regulated integrals in the range $2 < d < 4$;
integrating by parts we obtain a recursive equation for $k_n$, which
can be solved immediately, obtaining
$k_n = \Gamma(1-{d \over 2}-n)$, and consequently
\begin{equation}
b_n = {1 \over (4 \pi)^{d \over 2}} \, c_n(d) \,
\Gamma\!\left(1-{d \over 2}-n\right).
\label{eq:b_n}
\end{equation}
It is now trivial to set $d=3$ in Eq.\ (\ref{eq:c_n}) to obtain an
explicit recursive expression for $b_n$.

To compute numerical values of $a_n$ in $d=3$, it is not practical to
use Eq.\ (\ref{eq:a_n}), given the slow convergence of the integration
for large $\alpha$.  We found it more convenient to split the $\alpha$
integration in Eq.\ (\ref{eq:Xi-integral-repr}) at $\alpha=1$.  For
$\alpha\le1$ we can expand in powers of $m_0^2$ under the integration
sign and integrate term by term.  For $\alpha\ge1$, we subtract $n+l$
terms of the expansion (\ref{eq:I0-asympt}), expand the integrand in
powers of $m_0^2$ up to $O(m_0^{2n})$, and integrate term by term;
while the resulting integrals converge for any $l\ge0$, it is useful
to set $l\ge3$ to ensure fast convergence of the integration for large
$\alpha$.  The integral of the regulator can be computed analytically
in terms of the incomplete Gamma function, and the singular part of
its asymptotic expansion reproduces the singular part of the expansion
(\ref{eq:Xi-asympt}).

Collecting all numerical and analytical results, we obtain for $d=3$
the following asymptotic expansion:
\begin{eqnarray}
\Xi(m_0^2) &=& \beta_c - {1 \over 4\pi}m_0 
- 0.012164158583022\, m_0^2
+ {1 \over 32\, \pi}m_0^3 + 0.00083776240606293\, m_0^4 \nonumber \\
&-& {11 \over 2560\, \pi}m_0^5 - 0.000066743211781194\, m_0^6 
+ {281 \over 430080\, \pi}m_0^7 + 5.7884488124445 \times 10^{-6}m_0^8 
\nonumber \\
&-& {71 \over 655360\, \pi}m_0^9 - 5.320777180475\times 10^{-7}m_0^{10}
+ {7783 \over 403701760\, \pi} m_0^{11} \nonumber \\
&+& 5.107544566044 \times 10^{-8} m_0^{12}
- {70289 \over 19377684480\, \pi}m_0^{13} + O(m_0^{14}),
\end{eqnarray}
which gives an error smaller than $10^{-14}$ in the range $0 \leq
m_0^2 \leq 0.1$.

It is immediate to extract from this expansion the 3-$d$ large-$N$
nearest-neighbor model values
\[
\gamma_0 \approx -0.0060820792915113, \qquad
\delta_0 \approx 0.00083776240606293.
\]

The asymptotic expansion of $\Xi^{(n)}(m_0^2)$, the $n$th derivative
of $\Xi(m_0^2)$ with respect to $m_0^2$, is easily obtained from the
above expression. There is an obvious precision loss, but in the
above-mentioned range the error in the second derivative is still
smaller than $10^{-10}$.

In the following, we will need to compute $\Xi^{(n)}$ in a fast and
accurate way, for generic values of $m_0^2$; to this purpose, we
tabulated $\Xi$ for values of $m_0^2$ on a uniform grid with step
$h=10^{-3}$ and compute $\Xi^{(n)}(m_0^2)$ by $n{+}4$-point Lagrange
interpolation.

\section{Integral representations for the effective propagator}
\label{sec:int-eff-prop}

Let us now recall from Ref.\ \cite{CR} the following basic result:
\[
 D(k, m_0, 0) = {1 \over 2} \int_0^1 dx \int_{-\pi}^\pi 
{d^d q \over (2 \pi)^d} {1 \over [m_0^2 + 2\sum_\mu(1- z_\mu \cos q_\mu)]^2},
\]
where $z_\mu = \sqrt{1 -x(1-x)\hat k_\mu^2}$.

Using Schwinger representation we then obtain:
\begin{eqnarray}
 D(k, m_0, 0) &=& {1 \over 2} \int_0^1 dx  \int_0^\infty d \alpha\, 
\alpha\, e^{-\alpha m_0^2}\int_{-\pi}^\pi 
{d^d q \over (2 \pi)^d} e^{-2\alpha \sum_\mu(1- z_\mu \cos q_\mu)}
\nonumber \\
&=& {1 \over 2} \int_0^1 dx  \int_0^\infty d \alpha \,
\alpha\, e^{-\alpha m_0^2} \prod_\mu 
\Bigl[e^{-2 \alpha} I_0\bigl(2 \alpha z_\mu\bigr)\Bigr].
\label{eq:D-integral-repr}
\end{eqnarray}

Alternatively we might use the Schwinger representation directly and obtain
\[
 D(k, m_0, 0)= {1 \over 2} \int_0^\infty d s 
\int_0^\infty d t\, e^{-(s+t)m_0^2} \prod_\mu 
\left[e^{-2 (s+t)} I_0\bigl(2\sqrt{s^2+t^2+2 s\, t \cos k_\mu} \bigr)\right],
\]
that can be reduced to the previous one by the variable change 
$s = x\,\alpha$, $t= (1-x)\,\alpha$.

The direct numerical evaluation of Eq.\ (\ref{eq:D-integral-repr}),
and especially of its derivatives with respect to $m_0^2$, in $d=3$ is
difficult, particularly for small values of $k$ or $m_0$; the
convergence can be improved dramatically by adding and subtracting a
symmetric combination of Bessel functions:
\begin{eqnarray}
D(k, m_0, 0) &=& 
{1 \over 2} \int_0^1 dx  \int_0^\infty d \alpha \,
\alpha\, e^{-\alpha(6 + m_0^2)}
\Bigl[\prod_\mu I_0\bigl(2 \alpha z_\mu\bigr) -  
      I_0^3\bigl(2 \alpha \bar z\bigr) \Bigr] \nonumber \\
&-& {1 \over 2} \int_0^1 dx \,
{1\over{\bar z}^2}\,\Xi'\!\left({m_0^2+6\over{\bar z}} - 1\right),
\label{eq:D-subtr}
\end{eqnarray}
where $\bar z = {1\over3}(z_1+z_2+z_3)$; the subtracted integral and
its first few derivatives with respect to $m_0^2$ can now be computed
accurately by Gauss-Legendre integration on a small grid; the
integration in $x$ of $\Xi'$ and of its derivatives is also easy, once
a few singular terms of Eq.\ (\ref{eq:Xi-asympt}) have been
subtracted.

In order to identify the lattice contributions to the asymptotic
expansion of $D$ we must expand the integrand in powers of $m_0^2$.
The IR singularities turn into unsuppressed positive powers of
$\alpha$, present in the large-$\alpha$ regime when $z_\mu \to 1$.
These singularities become worse and worse with increasing powers of
$m_0^2$. In order to classify them according to their degree we need
to consider the homogeneous expansion of the integrand in powers of
$\alpha^{-1}$ and of $\sqrt{1-z_\mu^2} \equiv x(1-x)\hat k_\mu^2$.

Recalling once more the asymptotic expansion of the Bessel function,
we can write down the homogeneous expansion in the form:
\begin{equation}
\prod_\mu e^{-2 \alpha}I_0(2\alpha z_\mu) \approx 
{e^{-\alpha x(1-x)\hat k^2}\over (4 \pi \alpha)^{d \over 2}} 
\sum_{n=0}^\infty C_n [\,d, \alpha x(1-x) \hat k_\mu^2] \,\alpha^{-n},
\label{eq:homog}
\end{equation}
where in turn
\begin{equation}
 C_n [\,d, \alpha x(1-x) \hat k_\mu^2] \equiv 
\sum_{m=0}^{2n} \gamma_{nm}(d) [\alpha x(1-x) \hat k^2]^m
\label{eq:Cn}
\end{equation}
and $\gamma_{nm}(d)$ may show a dependence on the ratios 
${\hat k^{2p} / (\hat k^2)^p}$.

Repeating the procedure developed in the previous section we may now
obtain the following decomposition:
\begin{equation}
D(k,m_0,0) = \sum_{n=0}^\infty R_n(k) m_0^{2n} + D_s(k,m_0),
\label{eq:Ddec}
\end{equation}
where
\begin{equation}
R_n(k) \equiv {(-1)^n \over 2\,n!}\int_0^1 dx 
\int_0^\infty d\alpha\, \alpha^{n+1} 
    \Biggl[\prod_\mu e^{-2 \alpha}I_0(2\alpha z_\mu) -
        {e^{-\alpha x(1-x)\hat k^2}\over (4 \pi \alpha)^{d \over 2}}
        \sum_{m=0}^n C_m \,\alpha^{-m} \Biggr]
\label{eq:Rn}
\end{equation}
is the part of $A_n(k) \equiv R_n(k)+S_n(k)$ which shows a regular
dependence on $k$ in the $k \rightarrow 0$ limit, and
\begin{equation}
D_s(k,m_0) \equiv {1 \over 2}\int_0^1 dx \int_0^\infty d\alpha 
{e^{-\alpha x(1-x)\hat k^2}\over (4 \pi \alpha)^{d \over 2}} 
\sum_{n=0}^\infty 
\Biggl[e^{-\alpha m_0^2} - 
    \sum_{m=0}^{n-1}{(-\alpha m_0^2)^m \over m!} \Biggr]
C_n \,\alpha^{1-n}
\label{eq:Dsdec}
\end{equation}
is a calculable expression including all the singular dependence on
$k$ and admitting an asymptotic expansion in the form
\[
D_s(k,m_0) \approx \sum_{n=0}^\infty 
\bigl[ S_n(k) m_0^{2n}+ B_n(k) m_0^{2n+d-2} \bigr].
\]

In order to compute $D_s(k,m_0)$ let us notice that it may be also
expressed in the form
\begin{equation}
D_s(k,m_0) = \sum_{n=0}^\infty \sum_{m=0}^{2n} \gamma_{nm}(d) I_{nm}(m_0^2)
\label{eq:Ds1}
\end{equation}
where
\begin{equation}
I_{nm}(m_0^2) \equiv {1 \over 2}\int_0^1 dx 
[x(1-x)\hat k^2]^m \int_0^\infty d\alpha 
{e^{-\alpha x(1-x)\hat k^2}\over (4 \pi \alpha)^{d \over 2}} 
\Biggl[e^{-\alpha m_0^2} - 
    \sum_{m=0}^{n-1}{(-\alpha m_0^2)^m \over m!} \Biggr]
\alpha^{1+m-n}
\label{eq:Inm}
\end{equation}
enjoys the property
\[
{d I_{nm} \over dm_0^2} = - I_{n-1,m}.
\]

Therefore we only need to evaluate
\begin{eqnarray}
I_{0m} (m_0^2) &\equiv& {1 \over 2}\int_0^1 dx [x(1-x)\hat k^2]^m
\int_0^\infty d\alpha \, e^{-\alpha x(1-x)\hat k^2-\alpha m_0^2}\,
{\alpha^{1-{d \over 2}+m} \over (4 \pi)^{d \over 2}} \nonumber \\
&=& {1 \over 2}\,{\Gamma(2-{d \over 2}+m) \over (4 \pi)^{d \over 2}}
\int_0^1 dx
{[x(1-x)\hat k^2]^m \over [m_0^2+x(1-x)\hat k^2]^{2-{d \over 2}+m}}
\nonumber \\
&=& {1 \over 2}\,{\Gamma(2-{d \over 2}) \over (4 \pi)^{d \over 2}}
(-\hat k^2)^m\Biggl({\partial \over \partial \hat k^2}\Biggr)^m 
\int_0^1 {dx \over [m_0^2+x(1-x)\hat k^2]^{2-{d \over 2}}}.
\label{eq:I0m}
\end{eqnarray}

A trivial comparison with previous results shows that the last
integral is directly related to the continuum propagator by the
replacement $k^2 \rightarrow \hat k^2$, and therefore, setting $\hat
\xi^2 = 1 + 4 m_0^2 / \hat k^2$ we obtain
\begin{equation}
I_{0m}(m_0^2) = (-\hat k^2)^m
\Biggl({\partial \over \partial \hat k^2}\Biggr)^m 
\Biggl[a_0 \hat \xi^{d-3}(\hat k^2)^{{d \over 2}-2} +
    {b_0 \over \hat k^2 \hat \xi^2} \,
    {}_2F_1\!\Biggl({d-1 \over 2},1,{d \over 2},1-{1 \over \hat \xi^2}\Biggr)
        m_0^{d-2}\Biggr],
\label{eq:I0m1}
\end{equation}
where we must appreciate that the above expression is naturally
decomposed into an analytic and a nonanalytic term, which implies that
we can immediately relate all coefficients $S_n(k)$ to the derivatives
with respect to $\hat k^2$ of $\hat\xi^{d-3}(\hat k^2)^{{d \over 2}-2}$.

Straightforward manipulations lead to the general relationship
\begin{equation}
S_n(k) = {a_0 \over n!}(\hat k^2)^{{d\over 2}-2}
\Biggl({4 \over \hat k^2}\Biggr)^n \sum_{p=0}^n 
{(-1)^p\,\Gamma({d-1 \over 2}) \over \Gamma({d-1 \over 2}-n+p)}
\Biggl({\hat k^2 \over 4}\Biggr)^p \sum_{q=0}^{2p} 
{(-1)^q\, \Gamma({d \over 2}-1-n+p) \over\Gamma({d \over 2}-1-n+p-q)}
\, \gamma_{pq}(d).
\label{eq:Sn}
\end{equation}

The above expression brings into evidence a peculiar feature exhibited
by the functions $S_n(k)$ when $d=3$. In this case the arguments of
the function $\Gamma(1-n+p)$ appearing in the denominator are integer
nonpositive numbers whenever $p \not= n$, and therefore the
corresponding contributions to the sum vanish, giving
\begin{equation}
S_n(k) = (-1)^n {a_0 \over n!}(\hat k^2)^{-{1\over 2}} 
\sum_{q=0}^{2n} {(-1)^q\, \Gamma({1 \over 2}) \over\Gamma({1 \over 2}-q)}
\gamma_{nq}(3).
\label{eq:Sn1}
\end{equation}

When $d=3$ the singular contributions to $A_n(k)$ are proportional to
$(\hat k^2)^{-{1\over 2}}$ for all $n$, and this is basically a
consequence of the vanishing dependence on $\xi$ in the analytic part
of the continuum propagator.

The simplest example of this procedure is offered by the integral
representation of  $A_0(k,0)$:
\begin{equation}
A_0(k,0)= a_0\,(\hat k^2)^{{d \over 2}-2}+ R_0(k),
\label{eq:A0R0}
\end{equation}
where
\begin{equation}
R_0(k) \equiv {1 \over 2} \int_0^1 dx  \int_0^\infty d \alpha \,\alpha\, 
\Biggl[ \prod_\mu e^{-2 \alpha} I_0(2 \alpha z_\mu) 
    - {e^{-\alpha x(1-x)\hat k^2}\over (4 \pi \alpha)^{d \over 2}} \Biggr].
\label{eq:R0}
\end{equation}

This representation is worth a few observations. Notice that the
explicit term might be simply obtained from the expression of $A_{0c}$
by the replacement $k^2 \rightarrow \hat k^2$, and therefore it
reproduces exactly the singular behavior of $A_0$ when $k \rightarrow
0$. Moreover $R_0(k)$ has, by construction, a finite limit when $k
\rightarrow 0$. It is easy to check that $R_0(0)= - \gamma_0,$ thus
verifying explicitly that the choice $\gamma=\gamma_0$ leads to the
$O(a^{4-d})$ improvement of the lattice massless propagator.

By the same technique we obtain
\begin{equation}
A_1(k) = a_0 \Biggl[{2(d-3) \over \hat k^2}+{(6-d)(4-d) \over 16}
{\hat k^4 \over (\hat k^2)^2}-{(8-d) \over 16}\Biggr]
(\hat k^2)^{{d \over 2}-2}+ R_1(k),
\label{eq:A1R1}
\end{equation}
where
\begin{eqnarray}
R_1(k) &\equiv& -{1 \over 2} \int_0^1 dx  \int_0^\infty d \alpha \,\alpha^2 
\Biggl[ \prod_\mu e^{-2 \alpha} I_0(2 \alpha z_\mu) \nonumber \\
&-& {e^{-\alpha x(1-x)\hat k^2}\over (4 \pi \alpha)^{d \over 2}}
\Biggl(1+{d \over 16 \alpha} + {1 \over 4}x(1-x) \hat k^2 -
    {1 \over 4}\alpha x^2 (1-x)^2 \hat k^4 \Biggr) \Biggr].
\label{eq:R1}
\end{eqnarray}
Again we notice that the explicit term reproduces the leading
singularities, and moreover $R_1(0)= -\delta_0$.

We also mention the result for $A_2(k)$, since it will be needed in
the evaluation of $\delta g^{(1)}$:
\begin{equation}
A_2(k)= {1 \over 2} a_0 
\Biggl[ {a_{2,0} \over (\hat k^2)^2}
    + {a_{2,1} \over \hat k^2}+a_{2,2} \Biggr]
(\hat k^2)^{{d \over 2}-2}+ R_2(k),
\label{eq:A2R2}
\end{equation}
where
\begin{eqnarray*}
a_{2,0} &=& 4(d-3)(d-5), \\
a_{2,1} &=& {(d-3)(d-12)\over 8} +
{(d-3)(d-6)(d-8) \over 8}\,{\hat k^4 \over (\hat k^2)^2}, \\
a_{2,2} &=& {(d-16)(d-8) \over 512} +
{(d-4)(d-6)(d-8) \over 256}\,{\hat k^4 \over (\hat k^2)^2} \\
&+& {(d-4)(d-6)(d-8) \over 64}\,{\hat k^6 \over (\hat k^2)^3}
 + {(d-4)(d-6)(d-8)(d-10) \over 512}\,{(\hat k^4)^2 \over (\hat k^2)^4},
\end{eqnarray*}
and we obtain
\begin{equation}
R_2(k)\equiv {1 \over 4} \int_0^1 dx  \int_0^\infty d \alpha \,\alpha^3 
\Biggl[ \prod_\mu e^{-2 \alpha} I_0(2 \alpha z_\mu)
    - {e^{-\alpha x(1-x)\hat k^2}\over (4 \pi \alpha)^{d \over 2}}
    \bigl(1+ r_{2,1}(k)+r_{2,2}(k) \bigr) \Biggr],
\label{eq:R2}
\end{equation}
where in turn
\begin{eqnarray*}
r_{2,1}(k) &=& {d \over 16\,\alpha}+{1 \over 4}x(1-x)\hat k^2
- {1 \over 4} \alpha x^2 (1-x)^2 \hat k^4, \\
r_{2,2}(k) &=& {d(d+8) \over 512\,\alpha^2}
+ {1 \over 64\, \alpha}(d+2)x(1-x)\hat k^2+ {1 \over 64}\bigl(2 (\hat k^2)^2
+ (8-d)\hat k^4 \bigr)x^2(1-x)^2 \\
&-& {1 \over 16}\alpha x^3 (1-x)^3 (2 \hat k^6+ \hat k^4 \hat k^2)
+ {1 \over 32}\alpha^2 x^4 (1-x)^4 (\hat k^4)^2.
\end{eqnarray*}

An obvious relationship exists between the values $R_n(0)$ and the
coefficients of the analytic part of the asymptotic expansion of the
function $\Xi'(m_0^2)$. One easily finds that $R_n(0) = {1 \over
2} (n+1)a_{n+1}$.

It should be by now clear that a trivial generalization of the same
technique allows for an explicit, albeit more cumbersome, evaluation
of all the functions $B_n(k)$.
We could verify that the correct expressions for $B_0(k)$ and $B_1(k)$
are reproduced, and we computed $B_2(k)$ for a better accuracy check
of our numerical estimates.  The result is too cumbersome to be
reported here.

In practice, $A_0$, $A_1$, and $A_2$ can be computed numerically
in $d=3$ exploiting the subtraction (\ref{eq:D-subtr}); since
\begin{equation}
e^{-6t}\Biggl[\prod_\mu I_0(2 \alpha  z_\mu) - 
I_0^3(2 \alpha {\bar z})\Biggr] \approx
e^{-\alpha x(1-x)\hat k^2}\,x^2(1-x)^2\,
{3 \hat k^4 - \bigl(\hat k^2\bigr)^2 \over 48 (4\pi \alpha )^{3/2}},
\label{eq:subtracted-expansion}
\end{equation}
the subtraction is enough to regularize $A_1$ and $A_2$.  Let us
consider, e.g., $A_1$;
by straightforward manipulations we obtain
\begin{eqnarray}
A_1(k) &=& -{1\over2} \int_0^1 dx
\Biggl\{\int_0^1 d\alpha \, \alpha^2 \, e^{-6\alpha} 
\Biggl[\prod_\mu I_0(2 \alpha z_\mu) - I_0^3(2 \alpha{\bar z})\Biggr]
\nonumber \\ &&\qquad
+\,{1\over{\bar z}^3}\,\Xi''\!\left({6\over\bar z}-1\right)
- {1\over16\pi\bigl(x(1-x)\hat k^2\bigr)^{3/2}}
\Biggr\};
\label{eq:A1}
\end{eqnarray}
it is useful to overregulate the $x$ integration by subtracting the
r.h.s.\ of Eq.\ (\ref{eq:subtracted-expansion}), integrated over
$\alpha$.  For $A_2$ we follow the same procedure; the subtractions
are more complicated and not worth writing here.  The computation of
$A_0$ is similar but of course easier.

We verified explicitly that $A_n$, computed in this way, and $B_n$ are
consistent with Eq.\ (\ref{eq:D-expansion}) for $n\le2$.

\section{Numerical results}
\label{sec:numeric}

It is worthwhile to present a selection of the numerical results that
we obtained in $d=3$ for the nearest-neighbor formulation 
($c = -{1 \over 12}$).

\begin{figure}[tb]
\begin{center}
\leavevmode
\epsfig{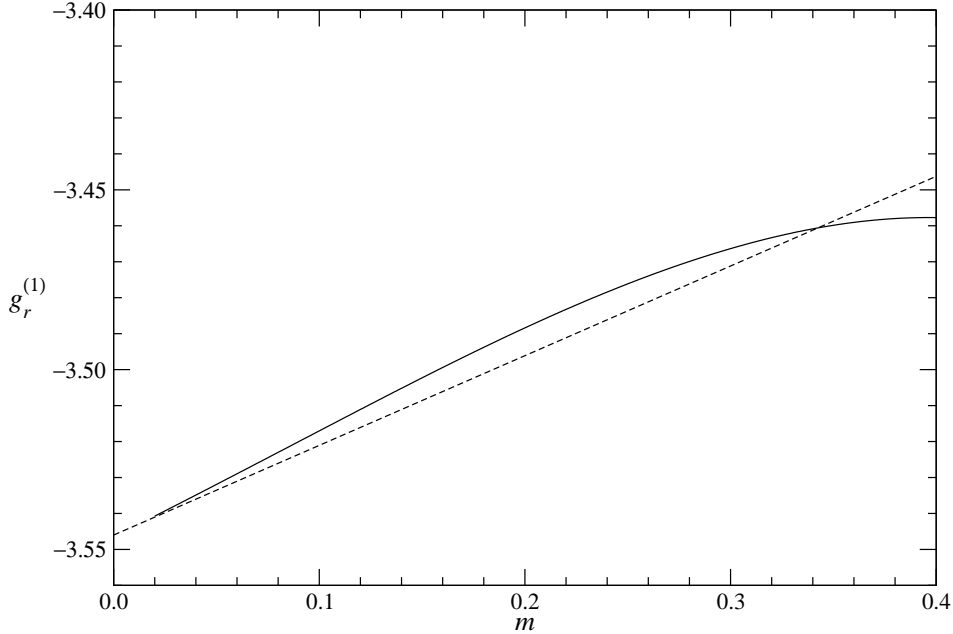}
\vskip-7mm
\end{center}
\caption{$g_r^{(1)}$ (solid line) and 
$g_{r,\rm con}^{(1)}+g_{r,\rm lat}^{(1)}$ (dashed line)
vs.\ $m$ for $\gamma_c=0$.
}
\label{fig:g1r-g0}
\end{figure}

\begin{figure}[tb]
\begin{center}
\leavevmode
\epsfig{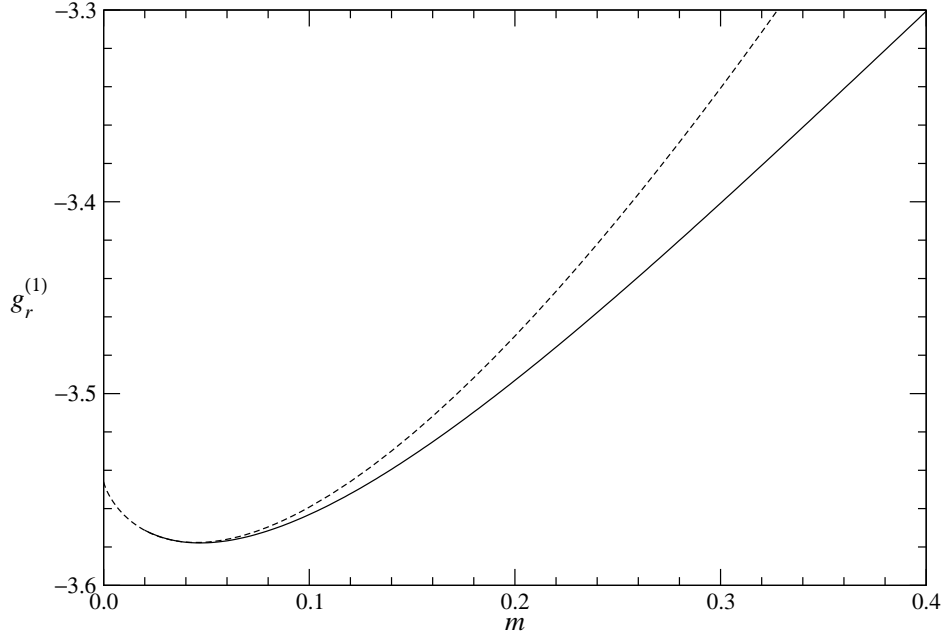}
\vskip-7mm
\end{center}
\caption{$g_r^{(1)}$ (solid line) and 
$g_{r,\rm con}^{(1)}+g_{r,\rm lat}^{(1)}$ (dashed line)
vs.\ $m$ for $\gamma_c=-\gamma_0$, i.e., for the nonlinear $\sigma$ model.
}
\label{fig:g1r-gg0}
\end{figure}

\begin{figure}[tb]
\begin{center}
\leavevmode
\epsfig{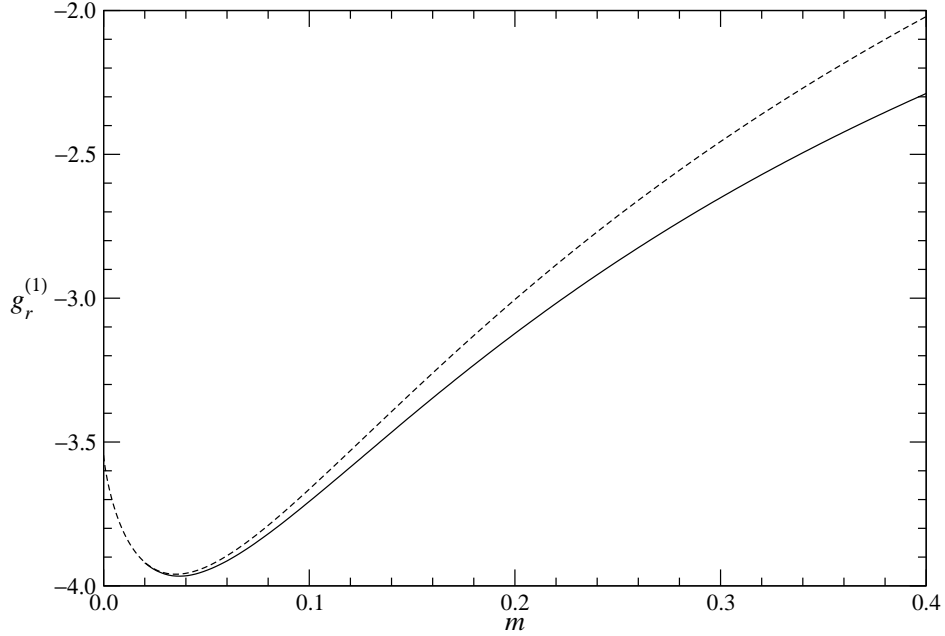}
\vskip-7mm
\end{center}
\caption{$g_r^{(1)}$ (solid line) and 
$g_{r,\rm con}^{(1)}+g_{r,\rm lat}^{(1)}$ (dashed line)
vs.\ $m$ for $\gamma_c=0.1$.
}
\label{fig:g1r-g0.1}
\end{figure}

In Figs.\ \ref{fig:g1r-g0}, \ref{fig:g1r-gg0}, and \ref{fig:g1r-g0.1}
we compare  $g_r^{(1)}(m, \gamma_c)$, obtained by direct evaluation of 
Eq.\ (\ref{eq:g1r-full}), with  
$g_{r,\rm con}^{(1)}(\gamma_c (ma)^{4-d}) + g_{r,\rm lat}^{(1)}(m, \gamma_c)$,
obtained from Eqs.\ (\ref{eq:g1r,con}), (\ref{eq:g1r,lat}), and
(\ref{eq:delta-g1r}), for three values of $\gamma_c$; of special
interest is the value $\gamma_c=-\gamma_0$, i.e., $\gamma=0$,
corresponding to the nonlinear $\sigma$ model.

\begin{figure}[tb]
\begin{center}
\leavevmode
\epsfig{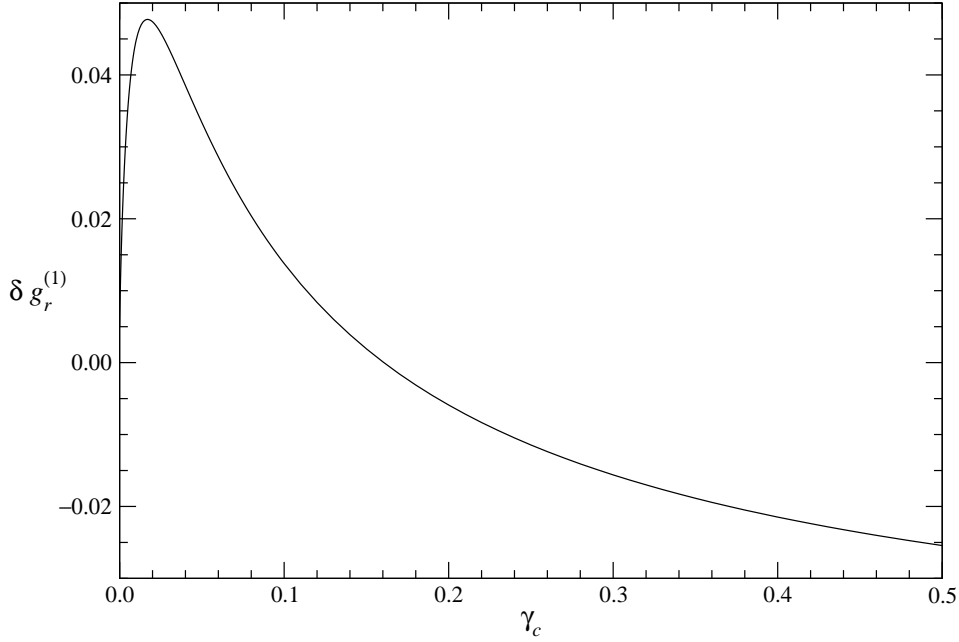}
\vskip-7mm
\end{center}
\caption{$\delta g^{(1)}$ vs.\ $\gamma_c$.
}
\label{fig:delta-g1r}
\end{figure}

$\delta g^{(1)}(\gamma_c)$ is plotted in Fig.\ \ref{fig:delta-g1r};
of special interest is the value
\[ \delta g^{(1)}(0) = 0.0049699... \]

\section{Construction of the improved Hamiltonian}
\label{sec:improved}

As we mentioned in the Introduction, the improvement procedure aims at
a systematic cancellation of the next-to-leading effects in the
invariant amplitudes.  The renormalization-group theory insures us
that a single choice of $\gamma$ exists such that this cancellation
occurs in all amplitudes.  It is therefore sufficient to find the
value $\gamma = \gamma^*$ for which the cancellation occurs in the
renormalized coupling.

In the context of the $1/N$ expansion we may assume $\gamma^*$ to
admit an expansion in powers of $1/N$:
\[
\gamma^* = \gamma_0^*+ {1 \over N} \gamma_1^* + O\!\left(1\over N^2\right).
\]

We have already recognized in Sec.\ \ref{sec:gap-eq} that $\gamma_0^*
= \gamma_0$.  We may therefore define 
\[
\gamma_c^* \equiv \gamma^* -\gamma_0 \approx {1 \over N} \gamma_1^*.
\]

Substituting this result in the expression of $g_r$ and expanding in
powers of $m$ we then obtain:
\[
g_r(m a, \gamma^*) \approx {1 \over d_0} 
\Biggl[1+{1 \over N}g_{r,\rm con}^{(1)}(0)
    + {1 \over N}{1 \over d_0}(\delta g_1(0)-\gamma_1^*)(m a)^{4-d}+O(m^2 a^2)
+ O\!\left(1\over N^2\right) \Biggr].
\]

We then recognize that the condition for the cancellation of the first
nonleading contribution is
\[
\gamma_1^* = \delta g_1 (0) \equiv 
\int_{-\pi}^\pi {d^d k \over (2 \pi)^d}
\Biggl({1 \over \tilde A_0}\Biggl(A_2-{\delta_0 \over \bar k^2}\Biggr)
    - {1 \over 2}\left({A_1 \over \tilde A_0}\right)^2 \Biggr)
+ \int_{-\infty}^{+\infty} {d^d k \over (2 \pi)^d } {4(d-3) \over (k^2)^2},
\]
where $\tilde A_0 \equiv A_0 (k, \gamma_0)$.

\section{Conclusions}
\label{sec:conclusions}

In the case of the three-dimensional $O(N)$ models with standard
nearest-neighbor interactions, our analytical results led us to the
prediction
\begin{eqnarray}
\gamma^* \cong -0.00608207 + {1\over N} \, 0.0049699 
+ O\!\left(1\over N^2\right).
\label{eq:gamma*}
\end{eqnarray}
We might now obtain an estimate for the value $N_c$ for which
$\gamma^*$ vanishes.  The numerical value of the present estimate is
admittedly not very promising, but, as we mentioned in the
introduction, the essential feature of $\gamma_1^*$ is its positive
sign, suggesting that $N_c$ may exist and possibly be within the range
of convergence of the $1/N$ expansion.

The physical interpretation of $N_c$ amounts to the statement that,
for this special value of $N$, the lattice version of the nonlinear
$\sigma$ model ($\gamma=0$) in three dimensions shows the absence of
leading irrelevant operators in the expansion of the Hamiltonian into
scaling fields.

For all values $N\le N_c$, one would get $\gamma^*>0$, and as a
consequence one may proceed to analyze the models in their improved
version, both by numerical Monte Carlo methods and by perturbative
expansion techniques.  Numerical evidence shows that this is actually
the case for the physically interesting cases $N\le3$.  It might be
interesting to perform a numerical study with the purpose of
estimating $N_c$.  A na\"\i ve extrapolation from the known numerical
values of $\gamma^*(N)$,
\[ \gamma^*(1) = 0.0159(3), \quad
   \gamma^*(2) = 0.0078(2), \quad
   \gamma^*(3) = 0.0043(4), \quad
   \gamma^*(4) = 0.0021(7),
 \]
obtained from Refs.\ \cite{CPRV-impr-O1}, \cite{CHPRV-impr-O2},
\cite{CHPRV-impr-O3}, and \cite{Hasenbusch-O4} respectively using the
formula $\gamma = \beta_c^2/(8\lambda N)$, suggests $N_c \simeq
5\div6$.

Concerning the extension of the $1/N$ expansion itself to the region
$N<N_c$, we must cautiously mention that some dramatic change in the
analytical behavior of the function $\gamma^*(N)$ may certainly occur
at $N=N_c$.  It is not possible to perform a strong-coupling expansion
of the models when $\gamma<0$, as one may immediately realize from an
analysis of the gap equation.

We would like to mention that the numerical evaluation of higher
orders of the $1/N$ expansion is technically not beyond reach, along
the lines traced by Ref.\ \cite{Flyvbjerg} and exploiting the more
accurate results for the effective propagator obtained in the present
paper.

Finally, as a consequence of the discussion of the previous sections,
it should be clear that, in the class of models with next-to-nearest
neighbor interactions, it is always possible to find a choice of
Hamiltonian parameters such that improvement becomes possible for
arbitrary values of $N$.

\end{document}